\documentclass[a4paper,fleqn]{cas-sc}
\usepackage{soul}
\usepackage[sort&compress,numbers]{natbib}
\usepackage{lineno}
\usepackage{setspace}
\usepackage{multirow}
\def\tsc#1{\csdef{#1}{\textsc{\lowercase{#1}}\xspace}}
\tsc{WGM}
\tsc{QE}
\tsc{EP}
\tsc{PMS}
\tsc{BEC}
\tsc{DE}
\begin{document}
\let\WriteBookmarks\relax
\def\floatpagepagefraction{1}
\def\textpagefraction{.001}
\shorttitle{Electronic, Optical, and Mechanical Properties of 2D C$_{60}$ Crystals}
\shortauthors{R. Tromer et~al.}
%\begin{frontmatter}

\title [mode = title]{A DFT Study of the Electronic, Optical, and Mechanical Properties of a Recently Synthesized Monolayer Fullerene Network}                
\author[1,2]{Raphael M. Tromer}
\author[3]{Luiz A. Ribeiro Junior}
\author[1,2]{Douglas S. Galv\~ao}[orcid=0000-0003-0145-8358]
\cormark[1]
\ead{galvao@ifi.unicamp.br}

\address[1]{Applied Physics Department, 'Gleb Wataghin' Institute of Physics, State University of Campinas, Campinas,SP, 13083-970, Brazil}
\address[2]{Center for Computing in Engineering \& Sciences, State University of Campinas, Campinas, SP, 13083-970, Brazil}
\address[3]{Institute of Physics, University of Bras\'{i}lia, Bras\'{i}lia, 70910-900, Brazil}

%\cortext[cor1]{Corresponding author}

\begin{abstract}
Closely packed quasi-hexagonal and quasi-tetragonal crystalline phase of C$_{60}$ molecules (named qHPC$_{60}$) was recently synthesized. Here, we used DFT simulations to investigate the electronic, optical, and mechanical properties of qHPC$_{60}$ monolayers. qHPC$_{60}$ has a moderate direct electronic bandgap, with anisotropic mechanical properties. Their elastic modulus ranges between 50 and 62 GPa. The results for optical properties suggest that qHPC$_{60}$ can act as UV collectors for photon energies until 5.5 eV since they present low reflectivity and refractive index greater than one. The estimated optical bandgap (1.5-1.6 eV) is in very good agreement with the experimental one (1.6 eV).
\end{abstract}

%\begin{graphicalabstract}
%\includegraphics[scale=0.08]{figs/graphical-abstract.png}
%\end{graphicalabstract}

%\begin{highlights}
%\item Highlight... ;

%\linenumbers


\maketitle
\doublespacing

\section{Introduction}

The advent of graphene in 2004 \cite{novoselov2004electric} has  stimulated the search for new carbon-based 2D materials that can share some of its remarkable properties \cite{kumar2018recent}, such as high mechanical strength, flexibility, thermal, and electrical conductivity \cite{abergel2010properties}. These unique properties made graphene one of the most promising materials for developing a wide range of applications \cite{rao2009graphene,westervelt2008graphene}. However, in spite of these remarkable properties, graphene is zero semi-metal bandgap, which limits its use in some electronics applications \cite{raccichini2015role}. For this reason, much effort has been dedicated in the search of new 2D carbon allotropes \cite{enyashin2011graphene,wang2015phagraphene,wang2018popgraphene,zhuo2020me,karaush2014dft,zhang2019art} that can present a non-zero bandgap while preserving some of the other graphene properties.

Novel 2D carbon-based materials with atomic thickness have emerged in the last two years through experimental investigations: the monolayer amorphous carbon (MAC) \cite{toh2020synthesis,Felix2020,Tromer2021} and the 2D biphenylene network (BPN) \cite{fan2021biphenylene}. Laser-assisted chemical vapor deposition was the synthesis route used to fabricate a free-standing, continuous, and stable MAC \cite{toh2020synthesis,Felix2020,Tromer2021}. This material possesses a pure carbon structure composed of randomly distributed five, six, seven, and eight atom rings. Its lattice arrangement differs from disordered graphene. BPN consists of the periodic lattice composed of a set of fused rings containing four, six, and eight carbon atoms. In the synthetic route used to produce BPN, an adsorbed halogenated terphenyl molecule undergoes a two-step interpolymer dehydrofluorination polymerization that creates the four- and eight-membered rings through carbon-carbon bond formation, employing a gold surface as substrate \cite{fan2021biphenylene}. Similar to graphene, MAC and BPN present a zero semi-metal bandgap.

Very recently, a single-crystal 2D carbon material with a semiconducting bandgap of about 1.6 eV, named monolayer quasi-hexagonal-phase fullerene (C60), was experimentally realized overcoming the problem of a null bandgap shown by other 2D carbon-based materials \cite{hou2022synthesis}. Through an organic cation slicing strategy, the exfoliation of quasi-hexagonal bulk single crystals yields the monolayer polymeric C60. The C$_{60}$ 2D crystals showed a large size via an interlayer bonding cleavage strategy. In these structures, C$_{60}$ polymers form covalently bonded cluster cages of C$_{60}$ in a plane. This clustering mechanism yielded two stable crystals of polymeric C$_{60}$ in closely packed quasi-hexagonal (qHPC$_{60}$) and quasi-tetragonal (qTPC$_{60}$) phases \cite{hou2022synthesis}. qHPC$_{60}$ and qTPC$_{60}$ exhibit high crystallinity and good thermodynamic stability. Their moderate bandgap and unique topological structure are desirable for developing new applications in nanoelectronics. Despite the insightful discussion on the physical properties of these 2D C$_{60}$ crystals presented in the original work \cite{hou2022synthesis}, a detailed description of their electronic, optical, and mechanical properties is still lacking.   

In this study, we numerically investigate the electronic, optical, and mechanical properties of qHPC$_{60}$ using density functional theory (DFT) methods. We restricted our study to the qHPC$_{60}$ phase since no monolayer phase was experimentally observed for the qTPC$_{60}$ one.  The calculated electronic and estimated optical bandgaps agree with the experimental findings. qHPC$_{60}$ has a direct bandgap with anisotropic mechanical properties. Consistent with the experimental findings the DFT results show that qHPC$_{60}$ is a 2D monolayer formed by covalent bonds of C$_{60}$ molecules along both directions of the crystal.      

\section{Methodology}

\begin{figure}[pos=t]
    \centering
    \includegraphics[scale=0.5]{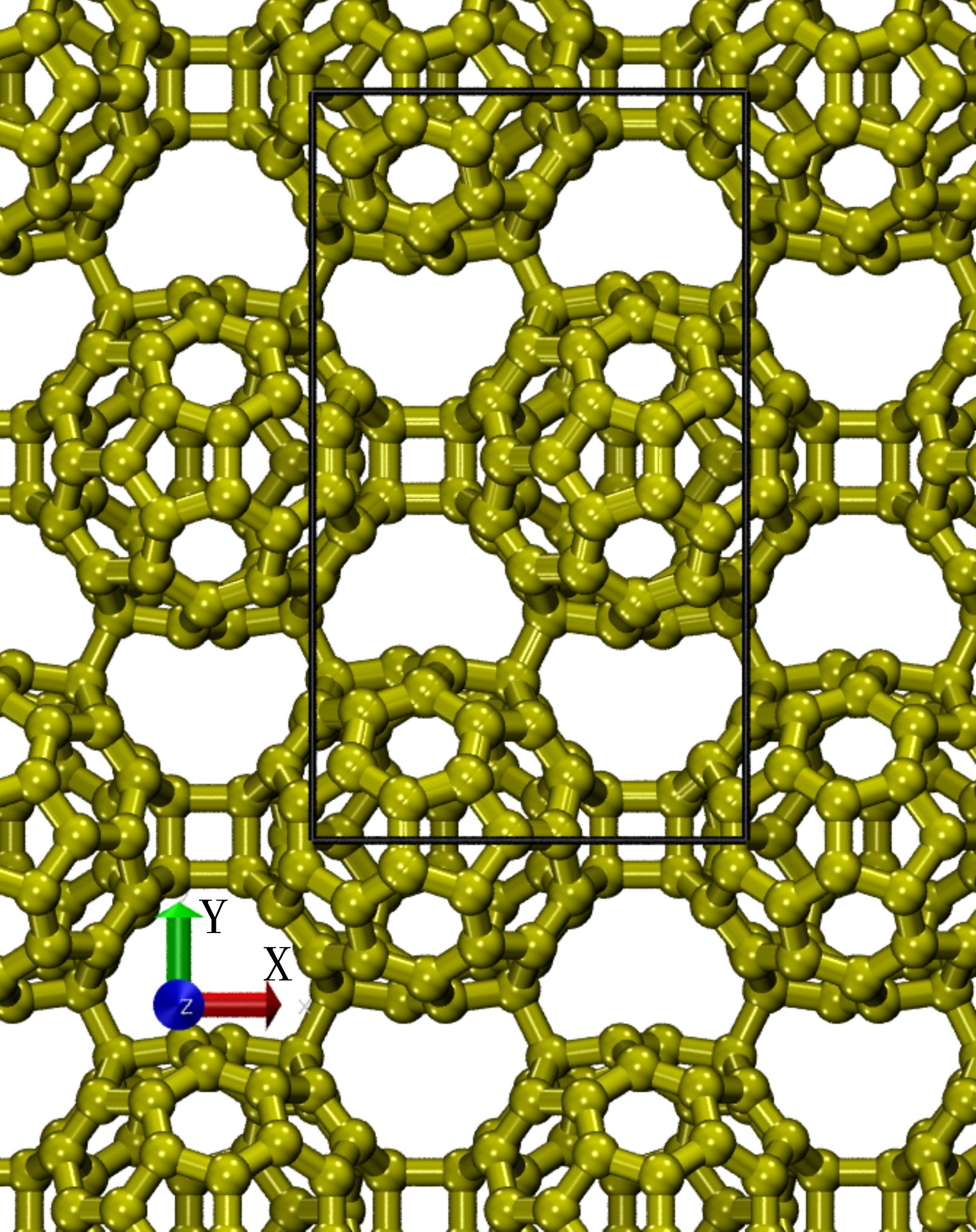}
    \caption{Schematic representation of the optimized atomic configuration for the qHPC$_60$ monolayers.}
    \label{fig1}
\end{figure}

The lattice structure of qHPC$_{60}$ is available from reference \cite{hou2022synthesis}. Its square unit cell consists of 120 carbon atoms, (see Figure \ref{fig1}). We fully optimized these structures and calculated their electronic, optical, and mechanical properties through DFT calculations, as implemented in the SIESTA code \cite{Soler2002,artacho2008siesta,artacho1999linear}. The van der Waals corrections were adopted to describe the exchange-correlation term \cite{vdw1,vdw2}. The calculations were carried out within the framework of the generalized gradient approximation (GGA) with the Perdew-Burke-Ernzerhof (PBE) functional \cite{perdew1996generalized,ernzerhof1999assessment} for the exchange-correlation term. We chose the norm-conserving Troullier-Martins pseudopotential to describe the core electrons \cite{troullier1991efficient}. The wave functions are based on localized atomic orbitals and a double-zeta plus polarization (DZP) basis set.   

We used 250 Ry for kinetic energy cut-off, sampling the reciprocal space with 20 $\times$ 20 $\times$ 2 k-point grid. The vacuum region along the z-direction is set to 30 \r{A} to prevent spurious interactions among layers. The basis vector along the z-direction was fixed during the optimization process. The other vectors and atoms were fully relaxed. The convergence criterion for maximum forces on each atom was 0.001 eV/\r{A}. Once the 2D C$_{60}$ crystals were fully optimized, we obtained the formation energy using the following expression: $E_{formation}=(E_{2D}-N_CE_{C})/N_C$, where E$_{2D}$ is the total energy of isolated 2D qHPC$_{60}$, E$_C$ is the energy of each isolated carbon atom, and N$_C$ is the total number of the carbon atoms.   

For the electronic band structure calculation, we considered the following path along the square lattice: $\Gamma = (0.0,0.0)$ to $X = (0.5,0.0)$ to $U = (0.5,0.5)$ to $Y=(0.0,0.5)$ to $\Gamma = (0.0,0.0)$. To obtain the bulk modulus from stress-strain relationship, we assumed the linear regime corresponding to 1\% percent of the total applied strain. 

To obtain the optical coefficients for each direction, we considered an external electric field of value 1.0 V/\r{A} \cite{Fadaie2016} applied along three possible polarization planes. The optical coefficients were extracted from complex dielectric function $\epsilon =\epsilon_1+i\epsilon_2$, in which the imaginary part, $\epsilon_2$, is derived from the direct interband transitions through Fermi's golden rule \cite{fermi},

\begin{equation}
\epsilon_2(\omega)=\frac{4\pi^2}{\Omega\omega^2}\displaystyle\sum_{i\in \mathrm{VB}, \, j\in \mathrm{CB}}\displaystyle\sum_{k} W_k \, |\rho_{ij}|^2 \, \delta	(\varepsilon_{kj}-\varepsilon_{ki}- \hbar \omega).
\label{e2}
\end{equation}

\noindent In Equation \ref {e2}, $\omega$ is the photon frequency, $VB$ and $CB$ are the valence and conduction bands, respectively, $\rho_{ij}$ is the dipole transition matrix element, $W_k$ is the weight of the respective k-point in the reciprocal space, and $\Omega$ is the unit cell volume.

\noindent The real part of the dielectric constant is obtained from the Kramers-Kronig relation \cite{kronig}: 

\begin{equation}
\epsilon_1(\omega)=1+\frac{1}{\pi}P\displaystyle\int_{0}^{\infty}d\omega'\frac{\omega'\epsilon_2(\omega')}{\omega'^2-\omega^2},
\end{equation}
\noindent where $P$ denotes the principal value of the logarithm function. Once the imaginary and real parts of the dielectric functions are obtained, one can derive, from simple relations, the other relevant properties, such as the absorption coefficient $\alpha$, the refractive index $\eta$, and the reflectivity $R$: 

\begin{equation}
\alpha (\omega )=\sqrt{2}\omega\bigg[(\epsilon_1^2(\omega)+\epsilon_2^2(\omega))^{1/2}-\epsilon_1(\omega)\bigg ]^{1/2},
\end{equation}
\begin{equation}
\eta(\omega)= \frac{1}{\sqrt{2}} \bigg [(\epsilon_1^2(\omega)+\epsilon_2^2(\omega))^{1/2}+\epsilon_1(\omega)\bigg ]^{2},
\end{equation}
\indent and,
\begin{equation}
R(\omega)=\bigg [\frac{(\epsilon_1(\omega)+i\epsilon_2(\omega))^{1/2}-1}{(\epsilon_1(\omega)+i\epsilon_2(\omega))^{1/2}+1}\bigg ]^2.
\end{equation}

It is well-known that in DFT methodologies, there is intrinsic imprecision in determining the bandgap values when contrasted to experimental results \cite{xiao2011accurate,crowley2016resolution}. This imprecision propagates in the calculation of optical coefficients. In the SIESTA code, this problem can be avoided from the optical properties by applying an ad-hoc shift of the unoccupied bands, known as the scissor operator. Here, we used a scissor equal to 0.6 eV.

\section{Results}

We begin our discussion by analyzing the optimized structure shown in Figure \ref{fig1}. After optimization, the for qHPC$_{60}$ lattice parameters are: a $=15.8794$ \r{A}, b $=9.1908$ \r{A}, c $=30.0000$ \r{A}, and $\alpha=\beta=\gamma=90^\circ$. The formation energy (E$_{formation}$) is -9.14 eV/atom, which is similar to those for other 2D carbon allotropes, like phagraphene \cite{Wang2015} and graphene \cite{tromer_2020}. It is worthwhile to stress that qHPC$_{60}$ is composed of sp$^3$ and sp$^2$ hybridized carbon atoms (see Supplementary Material). 

\begin{figure}[pos=t]
    \centering
    \includegraphics[scale=1.0]{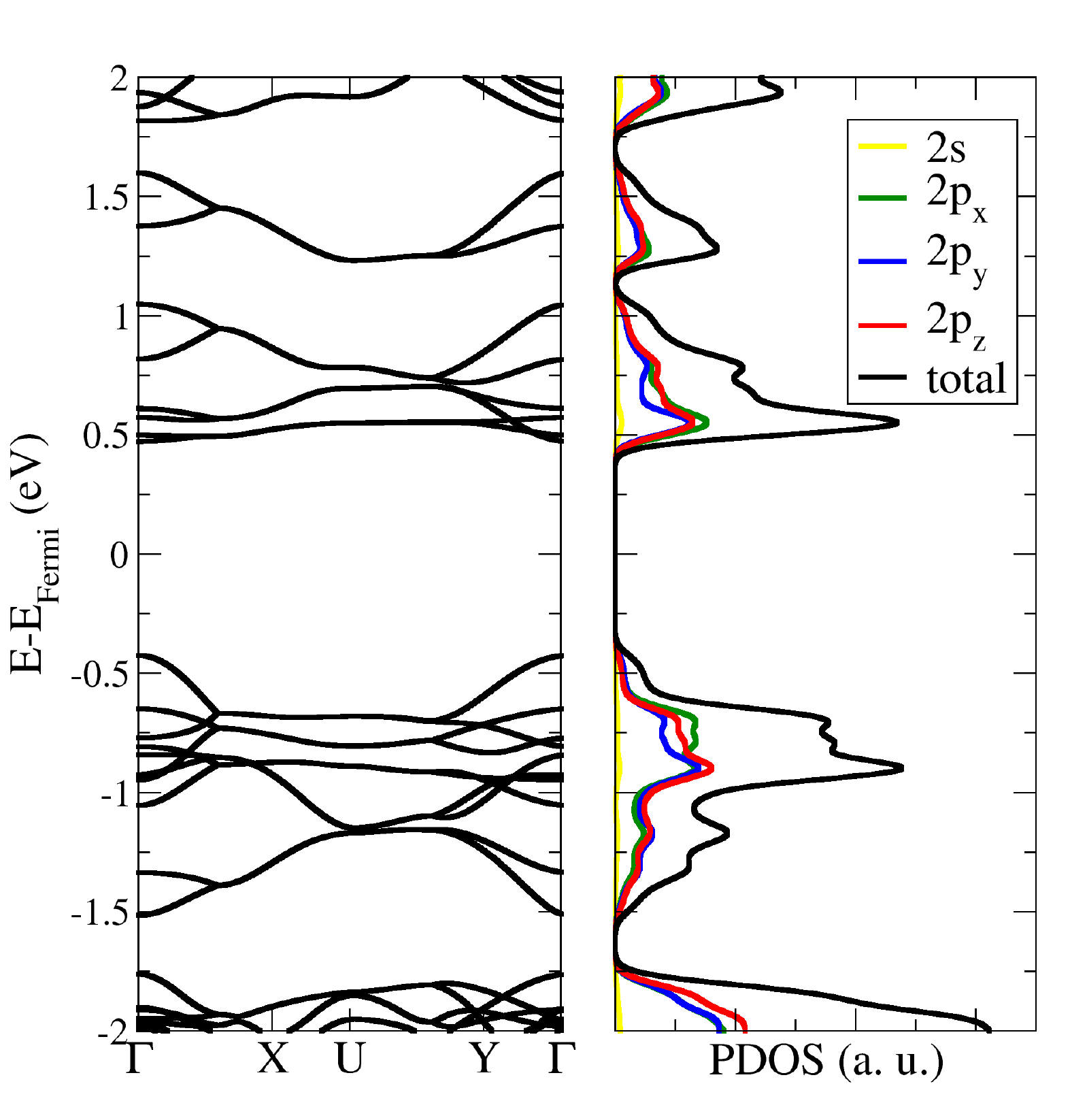}
    \caption{qHPC$_{60}$ electronic band structure and related projected density of states (PDOS), considering the valence atomic orbitals 2$s$ and 2$p$.}
    \label{fig2}
\end{figure}

In Figure \ref{fig2}, we present the qHPC$_{60}$ electronic band structure and related projected density of states (PDOS) considering the valence atomic orbitals 2$s$ and 2$p$. From the band structure, we can see that qHPC$_{60}$ is a small bandgap semiconductor. It has a direct bandgap at the symmetry point $\Gamma$ with a value of 0.9 eV, estimated from the energy difference between the lowest unoccupied molecular orbital (LUCO) and the highest occupied molecular orbital (HOCO). The PDOS show that HOCO and LUCO are predominantly composed of 2$p$ carbon atoms. We will discuss later the spatial distribution of these frontier orbitals.

\begin{figure}[pos=t]
    \centering
    \includegraphics[scale=0.51]{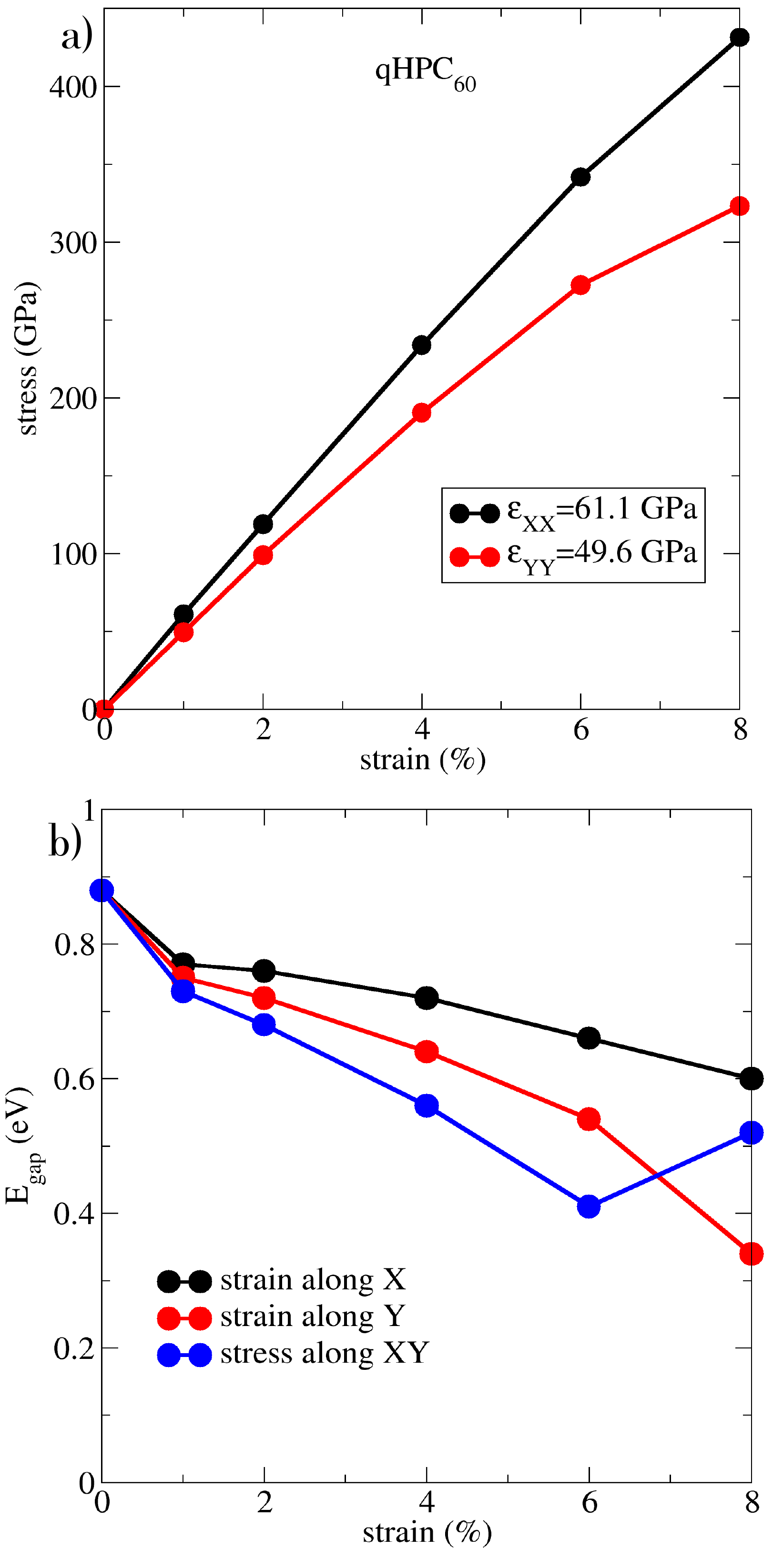}
    \caption{(a) Stress-strain curves for qHPC$_{60}$ monolayer for the x and y-directions. (b) Energy bandgap values as a function of the applied strain. Red, black, and blue refer to stress-strain interplay along the x, y, and x and y directions, respectively.}
    \label{fig3}
\end{figure}

As mentioned above, it is  well-known that standard DFT methods usually underestimate the bandgap values. For small bandgap semiconductors, the difference between computed and experimental bandgaps can reach 0.5 eV \cite{crowley2016resolution,xiao2011accurate}. It was reported in reference \cite{hou2022synthesis} that 2D C$_60$ crystal has a moderate bandgap of approximately 1.6 eV. Based on these facts, we can consider that the bandgap calculated here for qHPC$_{60}$ (0.9 eV) is in relative good agreement with the experimental value.   

To investigate the qHPC$_{60}$ mechanical properties, we applied a tensile strain (uniaxially and biaxially) and measured the tensile stress for the range of 0 to 8\% of strain. In Figures \ref{fig3}(a) and \ref{fig3}(b) we present the stress-strain curves under uniaxial strain loading along the $x$ and $y$ directions, respectively. To calculate Young's modulus values, ($\varepsilon_{XX}$ and $\varepsilon_{YY}$ along the x- and y-directions), we considered the linear regime until 1\% of strain. We can see from Figures \ref{fig3}(a) and \ref{fig3}(b) that there are two well-defined regimes for the stress-strain relationship. In the first regime, between 0 and 6\% of strain, the stress rises linearly. Above this critical threshold (second regime defined by 6 to 8 \% of strain) occurs a change in the slope of the line, and the stress increases slower when contrasted with the first regime. This trend for the stress-strain interplay suggests a structural phase transition.

The estimated elastic modulus were $\varepsilon_{XX}=49.6$ GPa and $\varepsilon_{XX}=61.1$ GPa. These results revealed that 2D C$_{60}$ crystals have clear in-plane anisotropic properties, as reported in the experiment \cite{hou2022synthesis} and expected from the structure topology. The square ring formed among the next-neighboring C$_60$ units (see Figure \ref{fig1}) results in a higher resilience to structural deformations along the y-direction.

\begin{figure}[pos=t]
    \centering
    \includegraphics[scale=0.45]{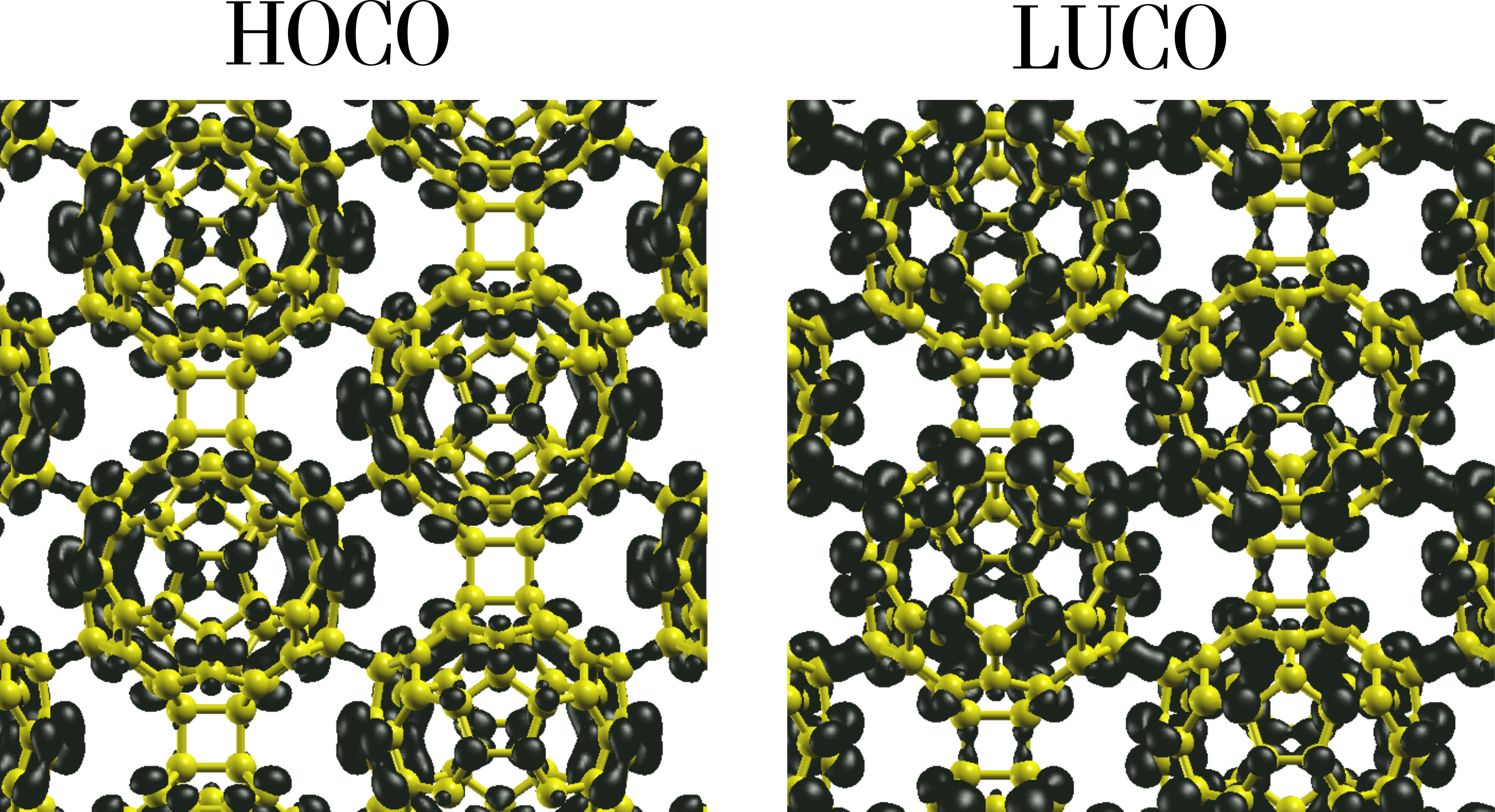}
    \caption{Schematic representation of the charge spatial distribution for the (left) HOCO and (right) LUCO energy bands. In the color scheme, black and yellow represent localized charges and carbon atoms/carbon-carbon bonds, respectively.}
    \label{fig4}
\end{figure}

We further explored the effects of strain engineering on the bandgaps values by applying uniaxial and biaxial strain in the range of 0 to 8\%. In Figure \ref{fig3}(b) we present the results of the calculated bandgaps as a function of the applied strain. We can see from the Figure that the bandgap values decrease linearly with increasing strain. As a general trend, the bond stretching process imposed by the applied tension increases the delocalization of the HOCO and LUCO orbitals decreasing the bandgap value. The exception is the bandgap increasing from 6\% to 8\% of strain in the biaxial case, in which the bandgap values increased from about 0.4 to 0.6 eV.  

As mentioned above, we calculated the HOCO and LUCO frontier crystalline orbitals. A schematic representation of their localization is illustrated in Figure \ref{fig4}.  We can see that HOCO and LUCO are well-spread over the carbon-carbon bonds, forming an electron transfer channel that can lead to high electron mobility. As mentioned above, the electronic states of HOCO and LUCO originate mainly from 2$p$ electrons. This HOCO and LUCO localization are consistent with the understanding that the higher the charge delocalization, the smaller the bandgap value. 

\begin{figure}
    \centering
    \includegraphics[scale=1.0]{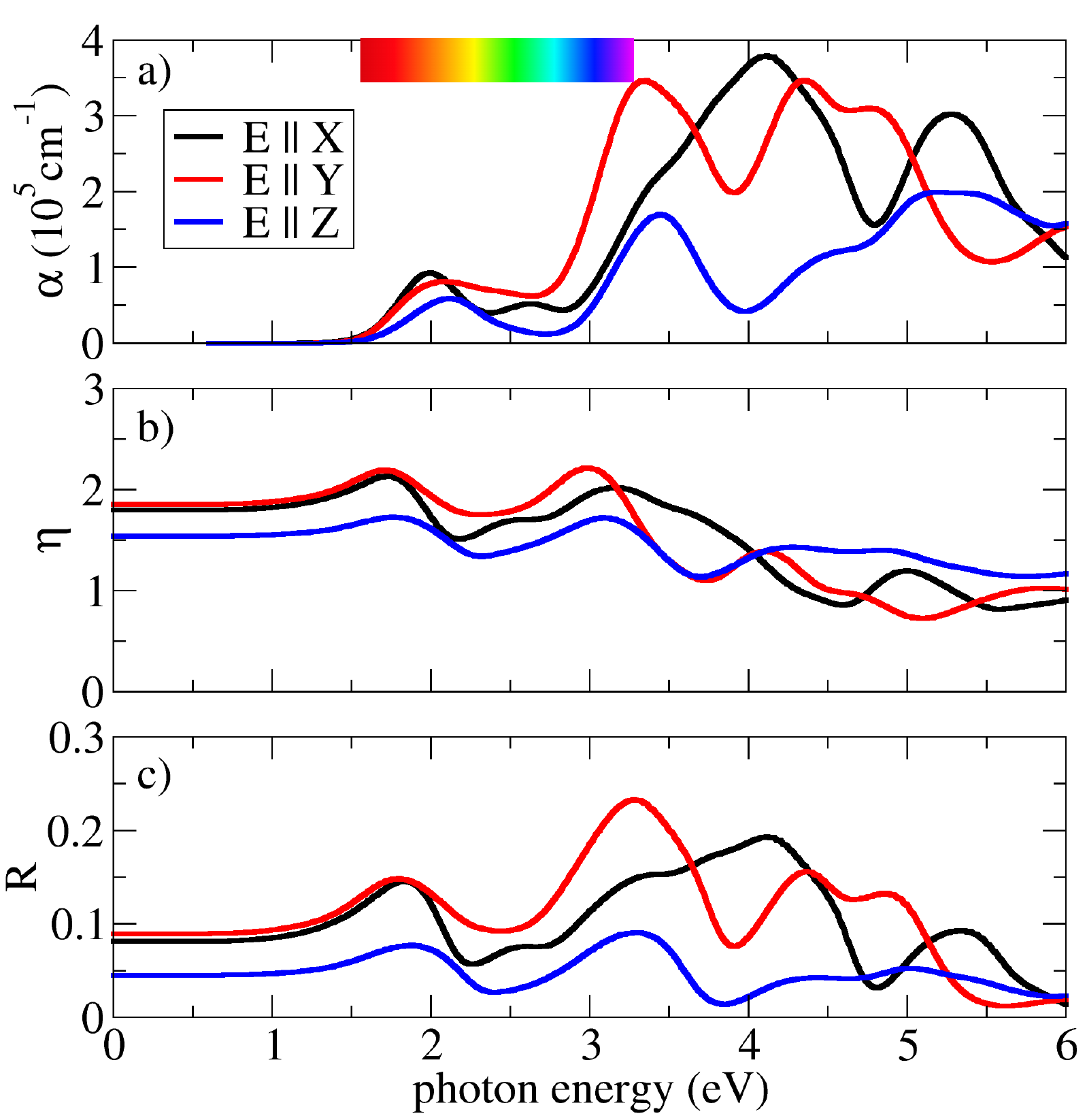}
    \caption{qHPC$_{60}$ absorption coefficient ($\alpha$), refractive ($\eta$), and reflectivity ($R$) indexes as a function of the photon energy value. Black, red, and blue lines refer to X, Y, and Z polarizations.}
    \label{fig5}
\end{figure}
 
Finally, we discuss the optical properties of the 2D C$_{60}$ crystal studied here. As mentioned in the Methodology section, the optical calculations considered the scissor correction to obtain a more accurate description of the optical transitions. Figures \ref{fig5}(a-c) show the absorption coefficient ($\alpha$), refractive ($\eta$), and reflectivity ($R$) indexes as a function of the photon energy value. From Figure \ref{fig5}(a), we can see that photon absorption begins approximately at 1.5-1.6 eV, which is in very good agreement with the experimental result, in which the optical bandgap value was also estimated to be at 1.55 eV \cite{hou2022synthesis}.

As a general feature, we observe a clear anisotropy among the three directions of the light propagation. For the $\alpha$ profile shown in Figure \ref{fig5}(a), we can see that the first optical transition peak occurs for nearly the same photon energy in all directions, about 2.2 eV (i.e., within the visible spectrum). This optical transition refers to the HOCO-LUCO transition. The other optical transitions occur within the violet and ultra-violet (UV) regions. The maximum absorption intensities are approximately $3.6 \times 10^5$ cm$^{-1}$. In the context of carbon-based materials, intense absorption within the UV region denotes the presence of sp$^{3}$-like carbons (see Supplementary Material).
 
Figures \ref{fig5}(b) and \ref{fig5}(e) show the refractive index as a function of the photon energy for X, Y, and Z polarizations. The maximum peak occurs at 1.8 eV (E$||$X). Light attenuation occurs for photon energies higher than 3.2 eV. $\eta$ values converge to 1.0, indicating that the incident UV light tends to be refracted similarly in all directions. The refractive index exhibits the same trend for photon energies between 3.2 and 6.0 eV.  In Figures \ref{fig5}(c) we can see that the reflectivity has the maximum peak at 3.3 eV (E$||$X, violet region) and 3.5 eV (E$||$Y, UV region), respectively. In both cases, almost 30\% of the incident light is reflected, and the other reflectivity peaks occur in the UV region. These results suggest that qHPC$_{60}$ can act as UV collectors for photon energies until 5.5 eV since they present low reflectivity and refractive index greater than one. 

\section{Conclusions}

In summary, we carried out DFT calculations to investigate the electronic, optical, and mechanical properties of the recently synthesized 2D C$_{60}$ crystals. In these materials, C$_{60}$ polymers form covalently bonded cluster cages of C$_{60}$ in a plane. This clustering mechanism yielded two stable structures named qHPC$_{60}$ and qTPC$_{60}$, but only qHPC$_{60}$ formed monolayers. From the electronic band structure, one can conclude that qHPC$_{60}$ is a small bandgap semiconductor with a direct bandgap with a value of 0.9 eV, consistent with the experimental data \cite{hou2022synthesis}. The estimated optical absorption bandgap is .5-1.6 eV, which is in very good agreement with the experimental result (1.55 ev).

Regarding the mechanical properties analysis, the calculated elastic modulus ranged from 50 to 62 GPa. The results also revealed that 2D C$_{60}$ crystals have clear in-plane anisotropic properties, as reported in the experiment \cite{hou2022synthesis} and expected from its topology with a higher mechanical resilience along the y-direction. The patterns of the calculated HOCO and LUCO frontier crystalline orbitals suggest a structure with high electron mobility.

The results obtained for the optical properties suggest that qHPC$_{60}$ can act as UV collectors for photon energies until 5.5 eV since it presents low reflectivity and refractive index greater than one. Moreover, our analysis points to the fact that these materials can serve as active layers in optoelectronic applications with photon energies varying from 1.5 to 5.5 eV, including the infrared and ultraviolet regions. 

%The supplementary material has additional data regarding electronic band structures and the calculation of the elastic moduli from molecular dynamics.

%\section*{Data availability statement}
%The data that support the findings of this study are available from the corresponding author upon %reasonable request.

\section*{Acknowledgements}
This work was financed in part by the Coordenação de Aperfeiçoamento de Pessoal de Nível Superior - Brasil (CAPES) - Finance Code 001, CNPq, and FAPESP. We thank the Center for Computing in Engineering and Sciences at Unicamp for financial support through the FAPESP/CEPID Grants \#2013/08293-7 and \#2018/11352-7. We also thank Conselho Nacional de Desenvolvimento Cientifico e Tecnológico (CNPq) for their financial support.  L.A.R.J acknowledges the financial support from a Brazilian Research Council FAP-DF grants $00193-00000857/2021-14$, $00193-00000853/2021-28$, and $00193-00000811/2021-97$ and CNPq grant $302236/2018-0$, respectively. L.A.R.J. gratefully acknowledges the support from ABIN grant 08/2019. L.A.R.J. acknowledges N\'ucleo de Computaç\~ao de Alto Desempenho (NACAD) for providing the computational facilities through the Lobo Carneiro supercomputer. L.A.R.J. thanks Fundaç\~ao de Apoio \`a Pesquisa (FUNAPE), Edital 02/2022 - Formul\'ario de Inscriç\~ao N.4, for the financial support. L.A.R.J acknowledges CENAPAD-SP for providing the computational facilities.

%% Loading bibliography style file
\bibliographystyle{unsrt}
%\bibliographystyle{model1-num-names}
%\bibliographystyle{cas-model2-names}

% Loading bibliography database
\bibliography{cas-refs}

\newpage

\section*{Supplementary Material}

\begin{figure}[pos=t]
    \centering
    \includegraphics[scale=0.8]{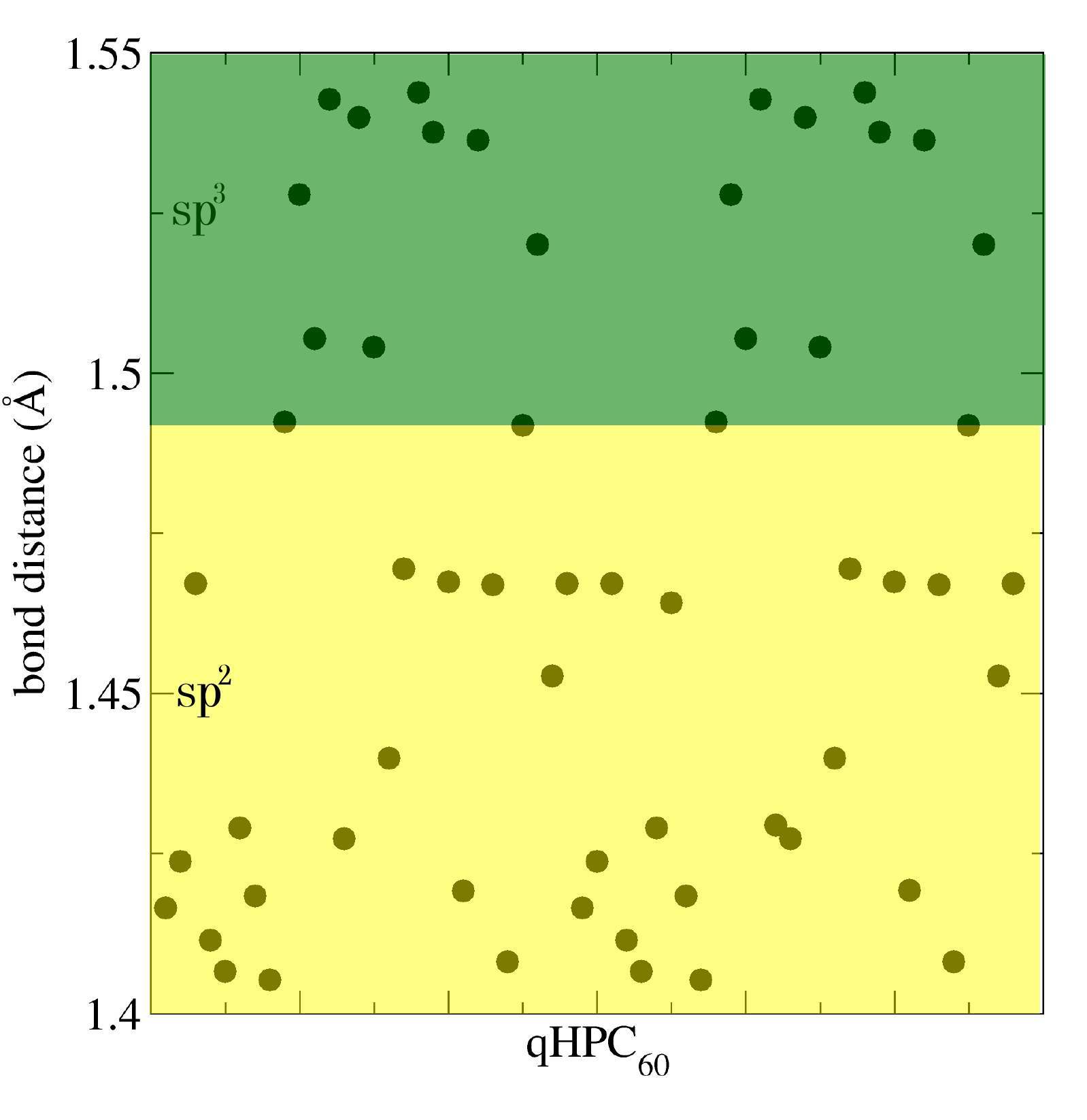}
    \caption{Bond distances within the unit cell for qHPC$_{60}$ and $qTPC_{60}$. We consider sp$^2$ hybridized carbon atoms the ones involved in carbon-carbon bonds with size until 1.48 \r{A}. sp$^3$ hybridized carbon atoms form bonds with size higher than this critical value. This bond configuration is in accordance with XPS measures reported in reference \cite{hou2022synthesis}.}
    \label{fig5}
\end{figure}

\end{document}